\documentclass[prb,twocolumn,byrevtex,floatfix,superscriptaddress]{revtex4}

\usepackage{amssymb,amsmath,graphicx,afterpage}
\usepackage{color,bm}
\pdfoutput=1

\begin{document}
\title{\boldmath N\'eel-type Skyrmion Lattice with Confined Orientation\\ in the Polar Magnetic Semiconductor GaV$_4$S$_8$ \unboldmath}
\author{I. K\'ezsm\'arki}
\affiliation{Department of Physics, Budapest University of
Technology and Economics and MTA-BME Lend\"ulet Magneto-optical
Spectroscopy Research Group, 1111 Budapest, Hungary}
\affiliation{Experimental Physics V, Center for Electronic
Correlations and Magnetism, University of Augsburg, 86135 Augsburg,
Germany}
\author{S. Bord\'acs}
\affiliation{Department of Physics, Budapest University of
Technology and Economics and MTA-BME Lend\"ulet Magneto-optical
Spectroscopy Research Group, 1111 Budapest, Hungary}
\author{P. Milde}
\affiliation{Institut f\"ur Angewandte Photophysik, TU Dresden,
D-01069 Dresden, Germany}
\author{E. Neuber}
\affiliation{Institut f\"ur Angewandte Photophysik, TU Dresden,
D-01069 Dresden, Germany}
\author{L. M. Eng}
\affiliation{Institut f\"ur Angewandte Photophysik, TU Dresden,
D-01069 Dresden, Germany}
\author{J. S. White}
\affiliation{Laboratory for Neutron Scattering and Imaging, Paul
Scherrer Institut, CH-5232 Villigen, Switzerland}
\author{H. M. R\o{}nnow}
\affiliation{Laboratory for Quantum Magnetism, \'Ecole Polytechnique
F\'ed\'erale de Lausanne, CH-1015 Lausanne, Switzerland}
\author{C. D. Dewhurst}
\affiliation{Institut Laue-Langevin, 6 rue Jules Horowitz, 38042
Grenoble, France}
\author{M. Mochizuki}
\affiliation{Department of Physics and Mathematics, Aoyama Gakuin
University, Sagamihara, Kanagawa 229-8558, Japan}
\affiliation{PRESTO, Japan Science and Technology Agency, Kawaguchi,
Saitama 332-0012, Japan}
\author{K. Yanai}
\affiliation{Department of Physics and Mathematics, Aoyama Gakuin
University, Sagamihara, Kanagawa 229-8558, Japan}
\author{H. Nakamura}
\affiliation{Department of Materials Science and Engineering, Kyoto
University, Kyoto 606-8501, Japan}
\author{D. Ehlers}
\affiliation{Experimental Physics V, Center for Electronic
Correlations and Magnetism, University of Augsburg, 86135 Augsburg,
Germany}
\author{V. Tsurkan}
\affiliation{Experimental Physics V, Center for Electronic
Correlations and Magnetism, University of Augsburg, 86135 Augsburg,
Germany}
\affiliation{Institute of Applied Physics, Academy of
Sciences of Moldova, MD 2028, Chisinau, Republica Moldova}
\author{A. Loidl}
\affiliation{Experimental Physics V, Center for Electronic
Correlations and Magnetism, University of Augsburg, 86135 Augsburg,
Germany}
\maketitle

\textbf{Following the early prediction of the skyrmion lattice
(SkL)~\cite{Bogdanov1989,Bogdanov1989_2,Bogdanov1994,Bogdanov1994_2,Rossler2006}---a
periodic array of spin vortices---it has been observed recently in
various magnetic crystals mostly with chiral
structure~\cite{Pfleiderer2004,Muhlbauer2009,Munzer2010,Yu2010,Yu2011,Wilhelm2011,Heinze2011,Seki2012,Adams2011,Adams2012,Tomomura2012,Yu2012,Romming2013,Milde2013,Shibata2013,Yu2014,Park2014,Janson2014}.
Although non-chiral but polar crystals with C$_{nv}$ symmetry were
identified as ideal SkL hosts in pioneering theoretical
studies~\cite{Bogdanov1989,Bogdanov1989_2,Bogdanov1994,Bogdanov1994_2},
this archetype of SkL has remained experimentally unexplored. Here,
we report the discovery of a SkL in the polar magnetic semiconductor
GaV$_4$S$_8$ with rhombohedral (C$_{3v}$) symmetry and easy axis
anisotropy. The SkL exists over an unusually broad temperature range
compared with other bulk
crystals~\cite{Pfleiderer2004,Muhlbauer2009,Munzer2010,Yu2011,Seki2012,Adams2012,Milde2013}
and the orientation of the vortices is not controlled by the
external magnetic field but instead confined to the magnetic easy
axis. Supporting theory attributes these unique features to a new
non-chiral or N\'eel-type of
SkL~\cite{Bogdanov1989,Bogdanov1994,Leonov2014} describable as a
superposition of spin cycloids in contrast to the Bloch-type SkL in
chiral magnets described in terms of spin
helices~\cite{Pfleiderer2004,Muhlbauer2009,Munzer2010,Yu2010,Yu2011,Wilhelm2011,Seki2012,Adams2012,Tomomura2012,Milde2013,Shibata2013,Park2014}.}

In non-centrosymmetric crystals, the energy associated with
ferromagnetic domain walls can be negative, hence, the homogeneous
ferromagnetic state becomes unstable against SkL
formation~\cite{Bogdanov1989,Bogdanov1989_2,Bogdanov1994,Bogdanov1994_2,Rossler2006,Yi2009}.
Two basic types of magnetic domain walls can form different
skyrmionic spin
textures~\cite{Bogdanov1989,Bogdanov1994,Bogdanov1994_2,Leonov2014}.
Bloch-type domain walls, where the spins rotate in the plane
parallel to the domain boundary, can form whirlpool-like skyrmions
with a given handedness as observed in chiral
magnets~\cite{Pfleiderer2004,Muhlbauer2009,Munzer2010,Yu2010,Yu2011,Wilhelm2011,Seki2012,Adams2012,Tomomura2012,Milde2013,Shibata2013,Park2014}.
In contrast, N\'eel-type domain walls, where the spins rotate in a
plane perpendicular to the domain boundary, can produce non-chiral
skyrmions. (See Fig.~5 for an illustration of the two skyrmion
types.) This second skyrmion archetype, expected to emerge in polar
magnets with C$_{nv}$ crystal
symmetry~\cite{Bogdanov1989,Bogdanov1994,Bogdanov1994_2,Leonov2014},
has not been observed yet. We reveal the formation of such a SkL in
GaV$_4$S$_8$, a magnetic semiconductor from this crystal symmetry
class, via magnetic susceptibility, atomic force microscopy (AFM)
and small-angle neutron scattering (SANS) measurements.

GaV$_4$S$_8$, a member of the lacunar spinel
family~\cite{Phouc2013,Elmeguid2004,Dorolti2010,Kim2014,Guiot2013,Singh2014,Pocha2000},
has a non-centrosymmetric cubic (T$_d$) structure at room
temperature~\cite{Pocha2000}. Lacunar describes the lack of every
second Ga atom compared to the normal spinel structure. These
ordered defects break the vanadium pyrochlore lattice into a network
of alternating larger and smaller V$_4$ tetrahedra, the so-called
breathing pyrochlore lattice. The magnetic building blocks are the
smaller V$_4$ clusters with spin 1/2, and form a face centered cubic
(FCC) lattice. The hybridization on a single V$_4$ unit leads to one
unpaired electron occupying a triply degenerate cluster orbital.
This orbital degeneracy is lifted by a cooperative Jahn-Teller
distortion that drives a cubic to rhombohedral structural transition
at $T_s$=42\,K~\cite{Pocha2000}. In the polar rhombohedral
(C$_{3v}$) phase the FCC lattice is stretched along one of the four
cubic $\langle$111$\rangle$ axes~\cite{Pocha2000}.

\begin{figure*}[t!]
\includegraphics[width=6.8in]{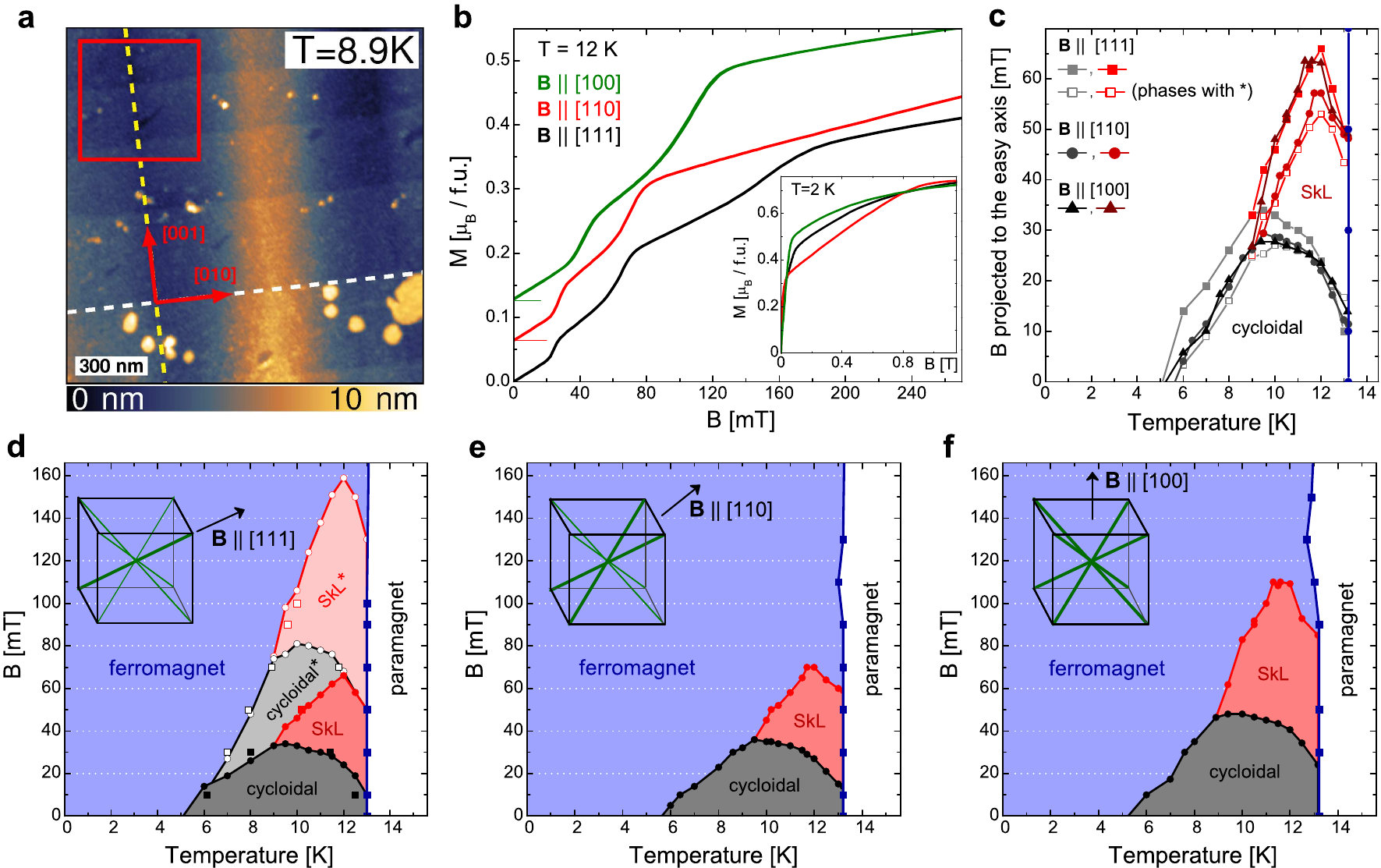}
\caption{\textbf{$\mid$ Magnetic phases in the lacunar spinel
GaV$_4$S$_8$.} \textbf{a,} Topographic image recorded by AFM on the
(100) surface of the crystal at $T$$=$8.9\,K. The colour scale
corresponds to the altitude perpendicular to the image plane.
Alternating blue and yellow stripes running along the cubic [001]
direction are structural domains with different rhombohedral axes,
while edges along the [010] axis are epilayer terraces. The yellow
and white dashed lines highlight a domain boundary and a step
between epilayer terraces, respectively. Magnetic patterns in Fig.~2
were recorded within the area of the red square. \textbf{b,}
Magnetization curves measured at 12\,K in
$\mathbf{B}$$\parallel$[100], [110] and [111] (shifted vertically
for clarity). Magnetization steps are observed at different field
values dependent on the orientation. The inset shows the
magnetization measured at $T$$=$2\,K up to $B$$=$1.2\,T.
\textbf{d-f,} Magnetic phase diagrams derived from the field- and
the temperature-dependence of the magnetization. Circles and squares
correspond to peaks in the field- and temperature-derivative of the
magnetization curves, respectively. The insets show the orientation
of the magnetic field relative to the easy axes of the four
rhombohedral domains (cubic body diagonals). The easy axes of
magnetically favoured/unfavoured domains are indicated by thick/thin
green lines. For $\mathbf{B}$$\parallel$[111], besides the cycloidal
and the SkL states, there are two additional phases extending up to
higher fields labelled as the cycloidal* and the SkL* states.
\textbf{c,} Phase boundaries from panels \textbf{d-f,} after
projecting the magnetic field onto the easy axis of the
corresponding domains.\label{fig0}}
\end{figure*}

The structural transition creates a multi-domain state with
submicron-thick sheets of the four different rhombohedral domains,
as seen in Fig.~1a. At $T_s$ the magnetic exchange interaction
changes from antiferro- to ferromagnetic and the material undergoes
a magnetic transition at $T_C$=13\,K, a temperature very close to
the Curie-Weiss temperature in the rhombohedral
phase~\cite{Yadav2008}. In the inset of Fig.~1b, the dependence of
the magnetization curves on the field orientation indicate a
considerable anisotropy. Indeed, our mean-field analysis shows the
ground state to be an easy axis ferromagnet with
$J_{\parallel}$$\approx$8.7\,K and $J_{\perp}$$\approx$8.3\,K as
described in the Supplementary Information. Here, $J_{\parallel}$
and $J_{\perp}$ respectively denote the exchange coupling for spin
components parallel and perpendicular to the easy axis, which itself
coincides with the direction of the rhombohedral distortion. These
exchange values are also consistent with $T_C$=13\,K.

GaV$_4$S$_8$ was considered as an ordinary ferromagnet below
$T_C$~\cite{Pocha2000,Yadav2008}, though there is a report about
low-field magnetic anomalies in this material~\cite{Nakamura2009}.
We found that the magnetic state is far from ordinary and contains
several neighbouring phases. Below $T_C$, the low-field
magnetization displays a sequence of steps indicating metamagnetic
transitions (see Fig.~1b). The positions of the steps, associated
with sharp peaks in the field-derivative of the magnetization, are
different for magnetic field, $\mathbf{B}$, applied parallel to the
[111], [110] and [100] cubic axes. The corresponding phase diagrams
are shown in Figs.~1d-f. Similar low-field steps have been observed
to separate the helical, SkL and conical states in
MnSi~\cite{Thessieu1997,Lamago2006,Muhlbauer2009},
Mn$_{1-x}$(Co,Fe)$_{x}$Si~\cite{Pfleiderer2010},
FeGe~\cite{Wilhelm2011} and Cu$_2$OSeO$_3$~\cite{Seki2012}. In
GaV$_4$S$_8$ we assign the three different magnetic phases as
cycloidal, SkL and ferromagnetic based on AFM imaging, SANS
experiments and theoretical calculations as described below.
\begin{figure*}[ht!]
\includegraphics[width=6.7in]{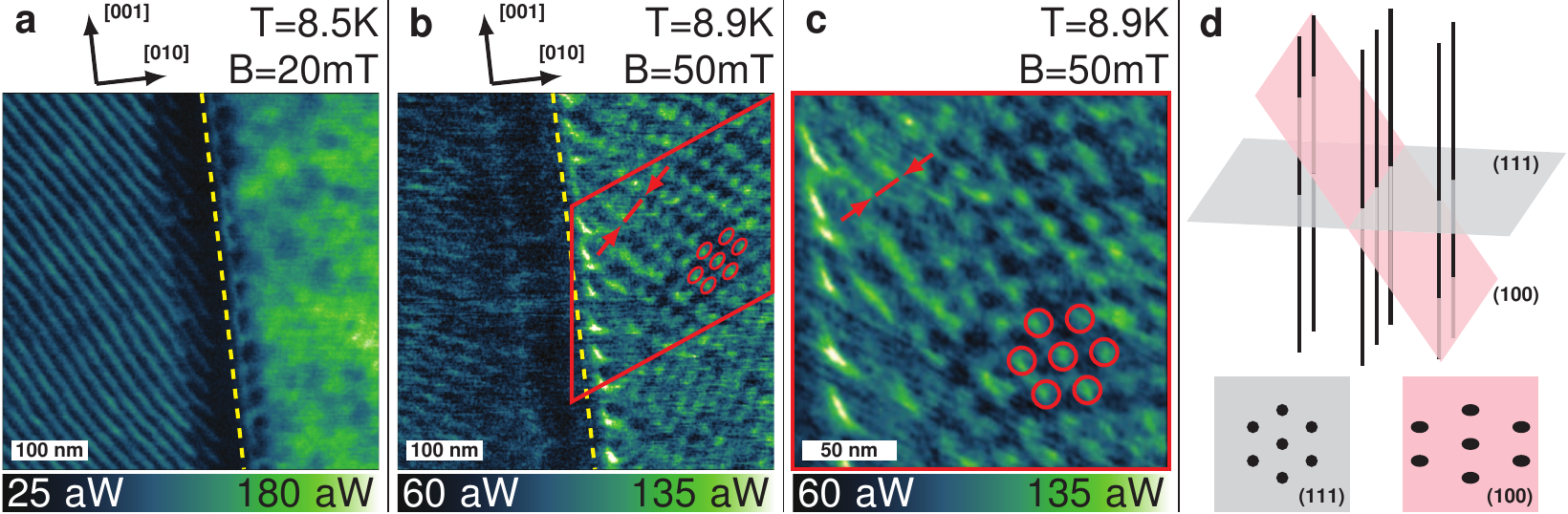}
\caption{\textbf{$\mid$ Real-space imaging of the magnetic patterns
in GaV$_4$S$_8$.} \textbf{a,} AFM image recorded at $T$$=$8.5\,K and
in $B$$=$20\,mT on the (100) surface in the area indicated by the
red square in Fig.~1a. Colour coding corresponds to dissipated power
due to magnetic interactions between the tip and the sample. A
modulated structure with a single $\mathbf{q}$-vector and a
corresponding periodicity of $a_{cyc}$$=$17.7$\pm$0.4\,nm is
observed on the left side of the domain boundary (yellow dashed
line). A similar pattern rotated by about 90 degrees emerges on the
other domain, though the magnetic contrast is weaker. \textbf{b,}
Magnetic pattern measured in the same area at $T$$=$8.9\,K and in
$B$$=$50\,mT. Distorted triangular lattices, rotated relative to
each other by approximately 90 degrees, are observed in both domains
with better contrast on the right side. The area with red border is
scaled by $\sqrt{3}$ according to the red arrows, i.e. perpendicular
to the direction of the shortest lattice periodicity. The area,
transformed to a square shape after this contraction, is displayed
in panel \textbf{c}. The green spots indicate a regular triangular
SkL with a lattice constant of $a_{sky}$$=$22.2$\pm$1\,nm and vortex
cores parallel to the [111] axis. For more details about the data
analysis and the evolution of the magnetic pattern as a function of
magnetic field see the Supplementary Information. \textbf{d,} Two
sections of a hexagonal array of vortex lines running along the
[111] axis. The intersections of the vortices with the (111) and
(100) planes respectively form a regular and distorted triangular
lattice. \label{fig0}}
\end{figure*}

For $\mathbf{B}$$\parallel$[111] there exist two additional magnetic
states besides the cycloidal, SkL and ferromagnetic states. This
arises due to multi-domain nature of the crystal and the easy axis
anisotropy, since only one of the domains has its easy axis parallel
to $\mathbf{B}$, while the easy axes of the other three
domains---[11$\bar{1}$], [1$\bar{1}$1] and [$\bar{1}$11]---lie at 71
degrees to $\mathbf{B}$. For these latter three domains the
non-collinear spin structures survive to higher fields because the
magnetic anisotropy energy is much larger than the Zeeman energy,
and only the field component parallel to the easy axis of each
domain can influence its magnetic state. This leads to the two
additional phases persisting up to $B$$=$80\,mT and 160\,mT, marked
with asterisks in Fig.~1d, and assigned as the cycloidal and SkL
states common for these three domains. For
$\mathbf{B}$$\parallel$[110], the cycloidal and SkL phases common
for the two domains with easy axes [111] and [11$\bar{1}$] are
indicated in Fig.~1e. The other two domains with easy axes
perpendicular to $\mathbf{B}$ cannot contribute to the magnetization
is this field range. For $\mathbf{B}$$\parallel$[100], the easy axis
of each domain span the same angle to $\mathbf{B}$, hence all
domains share common phase boundaries as shown in Fig.~1f. Phase
boundaries for all field orientations can be scaled together by
projecting the field onto the easy axes of the corresponding domains
(see Fig.~1c). Surprisingly, the SkL state exists over an unusually
broad temperature range, down to $\sim$0.68$T_C$, compared with
other bulk crystals where this phase is stable only over a 2-3\%
range immediately below $T_C$~\cite{Muhlbauer2009,Seki2012,Yu2011}.

To observe directly the spin pattern in real space we performed AFM
imaging in magnetic fields applied perpendicular to the sample
surface using magnetic cantilever tips~\cite{Milde2013}. For
$\mathbf{B}$$\parallel$[100], Figs.~2a and b show images of the same
surface area taken at $T$$=$8.5\,K and 8.9\,K in $B$$=$20\,mT and
50\,mT, respectively. The trail running along the [001] axis and
dividing each image into two halves is a structural domain boundary.
\begin{figure*}[ht!]
\includegraphics[width=7in]{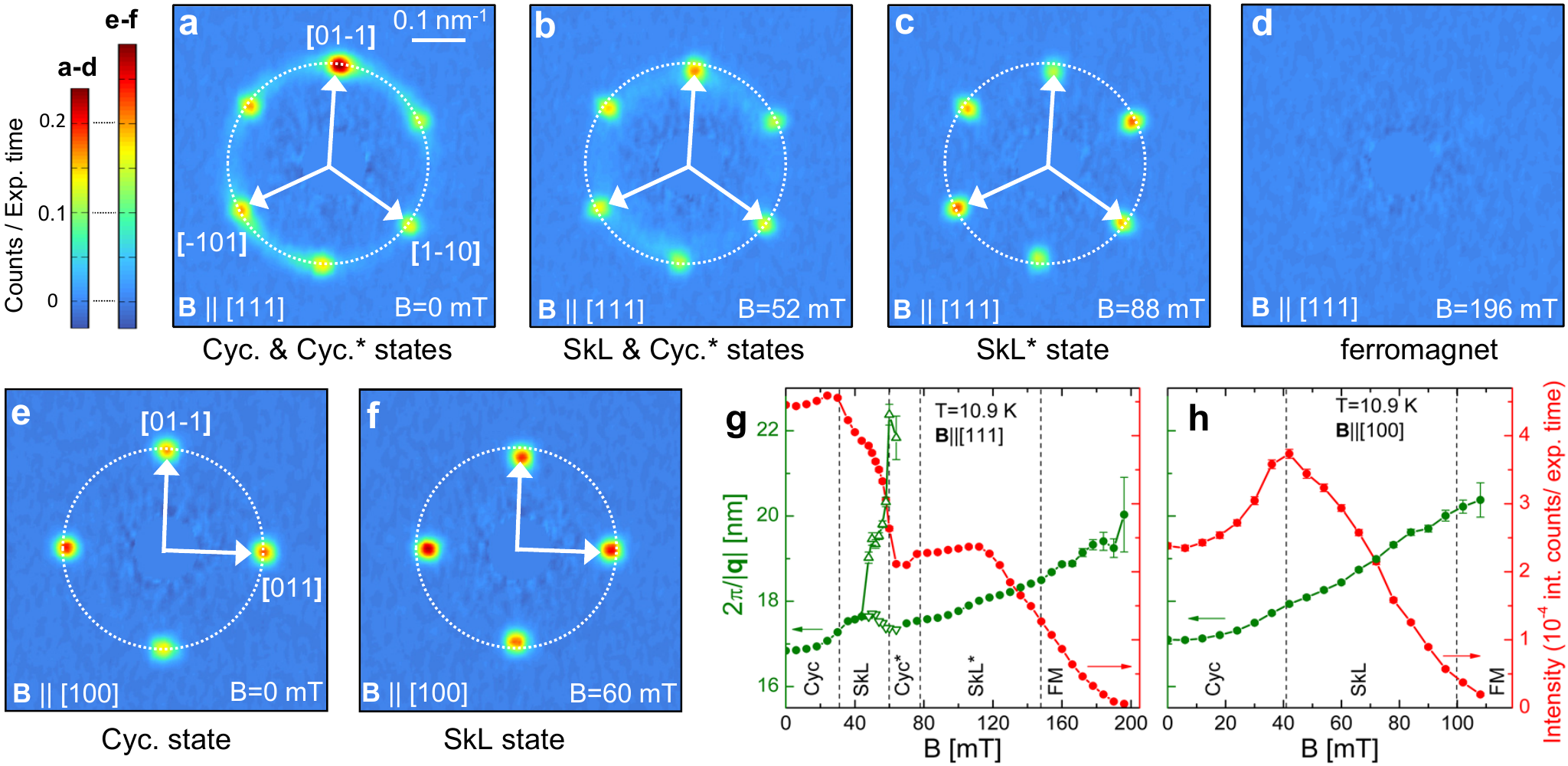}
\caption{\textbf{$\mid$ Small angle neutron scattering study of the
magnetic states in GaV$_4$S$_8$.} \textbf{a-d,} SANS images measured
at $T$$=$10.9\,K in various fields for $\mathbf{B}$$\parallel$[111].
The images were recorded at field values representative of the
different magnetic states assigned in Fig.~1d. \textbf{e \& f,} SANS
images taken at $T$$=$10.9\,K in different magnetic fields for
$\mathbf{B}$$\parallel$[100]. Labels refer to the magnetic states
assigned in Fig.~1f. The dashed circles having common diameter in
all the images help to visualize the change in the magnitude of the
$\mathbf{q}$-vectors. The two colour bars indicate the scattering
intensity for the two field orientations. \textbf{g \& h,} The
magnetic periodicity, $a=2\pi/|\mathbf{q}|$, (left scales) and the
total scattering intensity (right scales) as a function of the field
strength for the two magnetic field orientations. Vertical dashed
lines indicate the magnetic phase boundaries according to Fig.~1d
and f. The splitting of each Bragg peak into two peaks with
different $|\mathbf{q}|$ was resolved for
$\mathbf{B}$$\parallel$[111] in the field range of $B$$=$44-64\,mT
(see the Supplementary Information) which arises from the
coexistence of cycloidal* and SkL states in different
domains.\label{fig0}}
\end{figure*}

Although all four domains should host the same magnetic states for
$\mathbf{B}$$\parallel$[100], we found that the magnetic pattern
always changed at the domain boundaries, and often it was not
possible to achieve good contrast for the two sides simultaneously.
Since the wave vectors of the magnetic modulation
($\mathbf{q}$-vector) are expected to be perpendicular to the easy
axis, they can be different for different domains, resulting in a
rotation of the magnetic pattern between the two sides. In Fig.~2a,
a modulated structure with a single $\mathbf{q}$-vector,
representative of the zero- and low-field region, is observed with
good contrast on the left side. Our calculations described later
reproduce this phase as a cycloidal spin state, where the spins
rotate in a plane containing the $\mathbf{q}$-vector. This is in
contrast to the helical order generally found in other SkL
materials, where the spins rotate in the plane perpendicular to
$\mathbf{q}$. Assuming that the $\mathbf{q}$-vector is perpendicular
to the easy axis, which is either the [11$\bar{1}$] or [1$\bar{1}1$]
axis for the left-side domain, the periodicity of the cycloid in
$B$$=$20\,mT is $a_{cyc}$$=$17.7$\pm$0.4\,nm. (For more details see
the Supplementary Information.) This value is confirmed by our SANS
study.

With increasing field the cycloidal pattern transforms into a
distorted triangular lattice as shown in Fig.~2b. The same pattern
rotated by about 90 degrees appears on both sides of the domain
wall, though the contrast is weaker on the left side. As
demonstrated in Fig.~2c, re-scaling the lattice on the right side by
$\sqrt{3}$ perpendicular to the direction of the shortest lattice
constant leads to a regular triangular lattice with a lattice
constant of $a_{sky}$$=$22.2$\pm$1\,nm. This value is also
consistent with our SANS data. The observed distortion is naturally
explained by the orientational confinement of the SkL. Since the
vortex cores are not aligned parallel to the magnetic field but
instead the easy axis, which is either the [111] or [$\bar{1}11$]
axis for the right-side domain, they form a regular triangular
lattice in the hard plane. Therefore, the terminus of the vortices
on the (100) surface form a distorted triangular lattice as sketched
in Fig.~2d. When further increasing the field, the magnetic pattern
disappears indicating a uniform ferromagnetic state.

SANS was employed to confirm the bulk nature of the magnetic
structures and determine the $\mathbf{q}$-vectors directly. Sharp
magnetic Bragg peaks were always observed at $\mathbf{q}$-vectors
parallel to the cubic $\langle$110$\rangle$ directions irrespective
of the field orientation, as shown in Fig.~3. For
$\mathbf{B}$$\parallel$[111] the $\pm\mathbf{q}$-vectors form a
regular hexagon, while for $\mathbf{B}$$\parallel$[100] a square of
four spots is observed. These results, in agreement with the results
of AFM imaging, confirm the orientational confinement of the SkL,
i.e. the alignment of the vortex lines parallel to the easy axis of
each domain. It also shows that in both the cycloidal and SkL
states, the three possible $\pm\mathbf{q}$ pairs are parallel to the
three [110] axes lying in the hard plane. For
$\mathbf{B}$$\parallel$[111] a pale ring structure appears besides
the hexagon in low fields. This implies that the
$\mathbf{q}$-vectors are not fully pinned to the [110] axes but that
they can point in any direction within the hard plane, at least in
the surface and domain boundary regions observed by AFM.
\begin{figure*}[th!]
\includegraphics[width=6.4in]{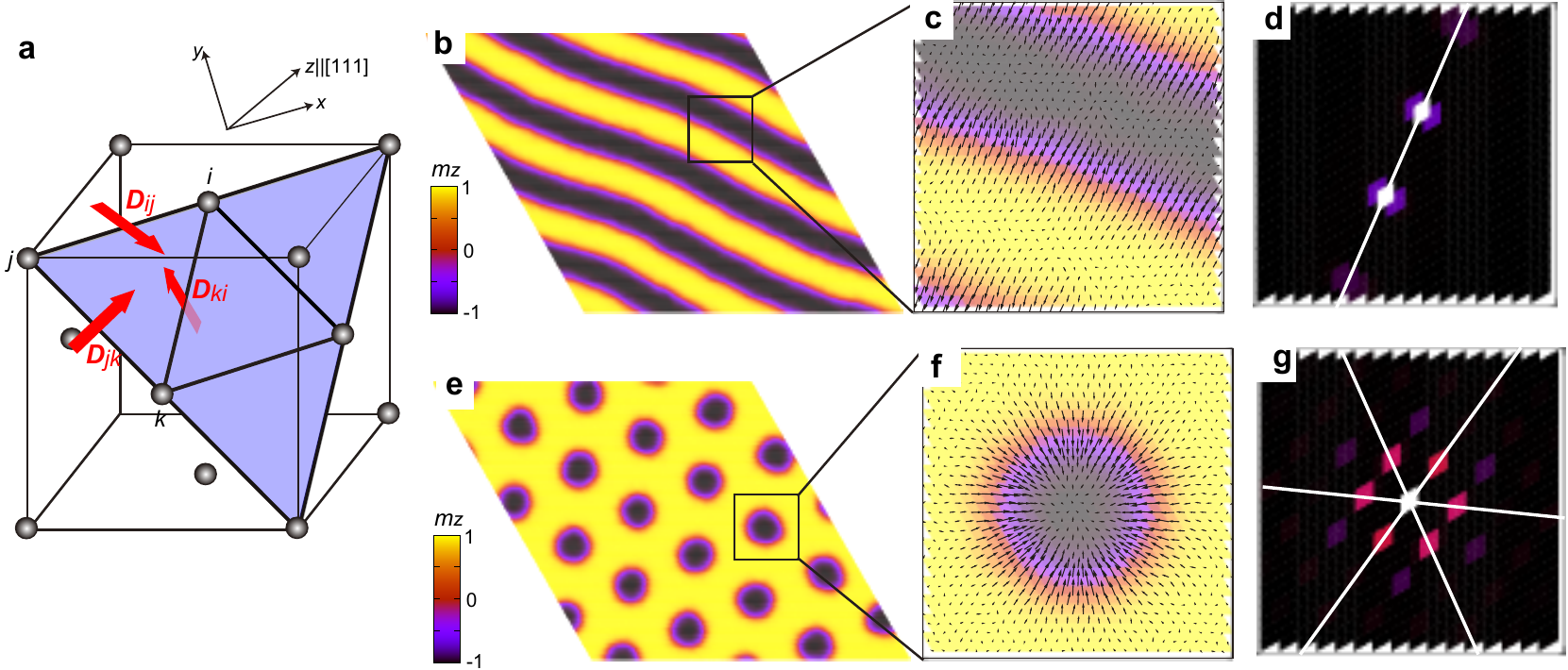}
\caption{\textbf{$\mid$ Spin patterns in the magnetic phases of
GaV$_4$S$_8$.} \textbf{a,} FCC lattice of V$_4$ units each carrying
a spin 1/2 and the orientation of the Dzyaloshinskii-Moriya vectors
for bonds on the triangular lattice within the (111) plane (chosen
as the $xy$ plane in the calculation). \textbf{b,} Cycloidal spin
state obtained for the spin model in Eq.~(\ref{eq:model}) on the
triangular lattice. \textbf{c,} Magnified view of the magnetization
configuration for the cycloidal state. \textbf{d,} Bragg peaks
($\mathbf{q}$-vectors) of the cycloidal state in panel \textbf{b} in
reciprocal space. \textbf{e,} SkL state obtained for the spin model
in Eq.~(\ref{eq:model}) on the triangular lattice. \textbf{f,} The
magnified view of the magnetization configuration for the SkL state
clearly shows the N\'eel-type domain wall alignment. \textbf{g,}
Bragg peaks of the SkL state in panel \textbf{e}. The
$\mathbf{q}$-vectors of the Bragg peaks are located along the [110]
directions (white lines) in the hard plane, for both the cycloidal
and SkL states.\label{Fig4}}
\end{figure*}

In each rhombohedral domain the cycloidal state can emerge with six
different $\mathbf{q}$-vectors, while the SkL state is unique, and
describable as a superposition of three cycloids whose
$\mathbf{q}$-vectors sum to zero. First-order Bragg peaks observed
using SANS do not allow a direct distinction between the cycloidal
and the SkL states due to the presence of magnetic and structural
domains. Nevertheless, the field dependence of the magnetic
periodicity ($a=2\pi/|\mathbf{q}|$) clearly indicates the phase
boundaries for both field orientations as shown in Fig.~3h. The
coexistence of the cycloidal* and SkL states in different domains
for $\mathbf{B}$$\parallel$[111] is also traced via the splitting of
$|\mathbf{q}|$.
\begin{figure}[t!]
\includegraphics[width=3.2in]{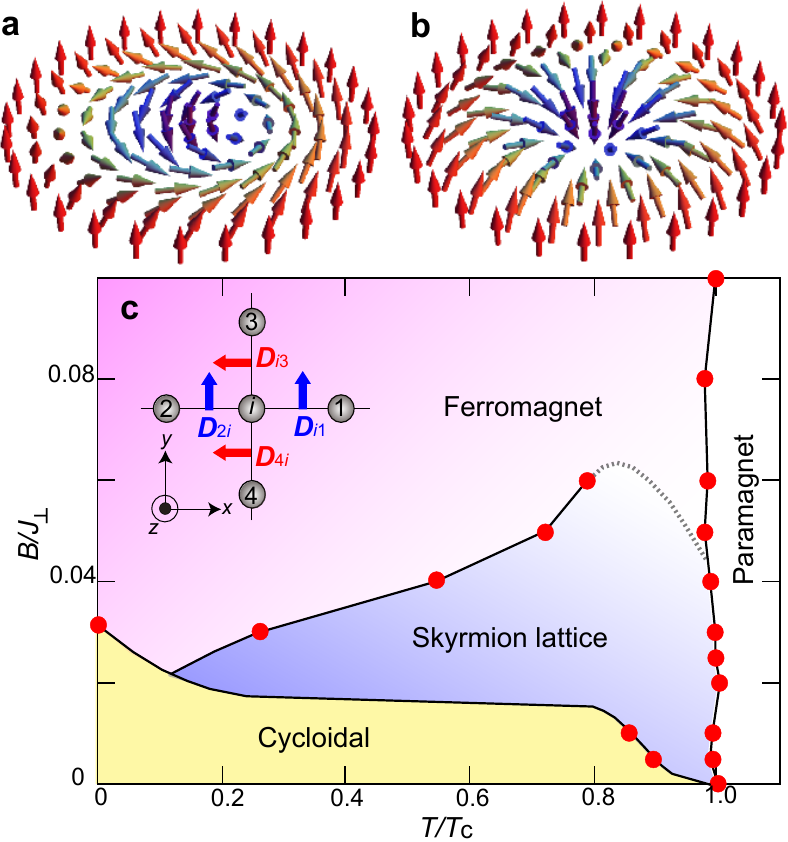}
\caption{\textbf{$\mid$ N\'eel-type SkL in GaV$_4$S$_8$.}
\textbf{a,} The Bloch-type skyrmion is a chiral object, i.e. it has
no mirror-plane symmetry. \textbf{b,} In contrast, the N\'eel-type
skyrmion has mirror-plane symmetry (combined with time reversal) for
vertical planes through its center. \textbf{c,} Theoretically
obtained temperature (T) versus magnetic field (B) phase diagram for
the spin model by Eq.~(\ref{eq:model}) on the cubic lattice. The
inset shows the direction of the Dzyaloshinskii-Moriya vectors for
bonds within the $xy$ plane. The skyrmion lattice state corresponds
to a N\'eel-type SkL.} \label{Fig5}
\end{figure}

In GaV$_4$S$_8$, the S$=$1/2 spins sit on an FCC lattice stretched
along the [111] direction, which can be regarded as triangular
lattices stacked along [111]. To determine the spin patterns in the
different phases we studied the following classical Heisenberg model
on the triangular lattice by a Monte Carlo technique;
\begin{eqnarray}
\mathcal{H}= \mathcal{H}_{\rm ex} + \mathcal{H}_{\rm DM} +
\mathcal{H}_{\rm Zeeman}, \label{eq:model}
\end{eqnarray}
where
\begin{eqnarray}
&&\mathcal{H}_{\rm ex}=-\sum_{<i,j>}\left(
 J_{\perp} m_{xi} m_{xj} + J_{\perp} m_{yi} m_{yj} +J_{\parallel} m_{zi} m_{zj} \right),\nonumber\\
&&\mathcal{H}_{\rm DM}
=\sum_{<i,j>}\bm D_{ij}(\bm m_i \times \bm m_{j}),\nonumber\\
&&\mathcal{H}_{\rm Zeeman}=-B \sum_{i}m_{zi}.\nonumber
\end{eqnarray}
For the exchange interaction $\mathcal{H}_{\rm ex}$, we consider
ferromagnetic coupling with XXZ-type anisotropy, where the $z$
(easy) axis is parallel to the [111] axis. The direction of the
Dzyaloshinskii-Moriya vectors, $\bm D_{ij}$, are determined by the
FCC crystal symmetry as shown in Fig.~\ref{Fig4}a. For the
calculation we chose $J_{\parallel}/J_{\perp}$=1.06 and
$D/J_{\perp}$=0.35, where $D=|\bm D_{ij}|$.

This model provides three ordered phases: the cycloidal, SkL and
ferromagnetic states. Figure~4 displays the spin patterns obtained
for the cycloidal and SkL states. The skyrmions exhibit the
N\'eel-type domain wall alignment along the radial direction from
their cores to their peripheries. Namely, the spins rotate within
the plane parallel to the radial direction. The Fourier components
of the magnetization configurations in the cycloidal and the SkL
states are displayed in Figs.~4d and g, respectively. These results,
in full agreement with SANS data, show that the
$\pm\mathbf{q}$-vectors are parallel to the [110] axes in the hard
plane in both states.

Extension of the model in Eq.~(\ref{eq:model}) to the distorted FCC
lattice requires additional parameters as both
$J_{\parallel}/J_{\perp}$ and $D/J_{\perp}$ can be different for
bonds within and between the triangular lattice planes. Instead, we
adapted the continuum model developed by Bogdanov and Hubert for
$C_{nv}$ symmetry~\cite{Bogdanov1994} to the cubic lattice in order
to reproduce the main features of the phase diagram. Here, $\bm
D_{ij}$s are perpendicular to the bonds along the $x$ and $y$
directions as shown in the inset of Fig.~5c, and values of
$J_{\parallel}/J_{\perp}$=1.06 and $D/J_{\perp}$=0.48 were used. We
stress that irrespective of the details of possible models,
GaV$_4$S$_8$ can only host a N\'eel-type SkL due to the orientation
of Dzyaloshinskii-Moriya vectors governed by its C$_{3v}$
symmetry~\cite{Bogdanov1989,Bogdanov1989_2,Bogdanov1994,Bogdanov1994_2,Leonov2014}.

Figure~5c displays the phase diagram obtained for the simple cubic
lattice. In agreement with experiment the triangular SkL appears
only at finite temperatures. The skyrmions have the same N\'eel-type
pattern, shown in Fig.~5b, as obtained for the triangular lattice
model. The cycloidal phase remains stable even at $T$=0 in contrast
to the experimental observation. This discrepancy likely arises from
not including corrections due to quantum fluctuations in our
classical Heisenberg model.

In a new class of materials we observed a SkL state emerging over an
extraordinarily broad temperature range. While chiral or Bloch-type
skyrmions have been investigated extensively in chiral magnets,
lacunar spinels provide a unique arena to study another topological
spin pattern, the non-chiral or N\'eel-type skyrmions. We found that
the strong orientational confinement of the vortices ensures the
robustness of two distinct skyrmionic states with a core
magnetization pointing either up or down the easy axis. This may
facilitate a unique magnetic control of the SkL, since we expect
that the SkL induced by a magnetic field parallel to the easy axis
can be rotated within the hard plane by an additional transverse
field component. Such magnetic control is inconceivable in cubic
helimagnets where vortex cores instantaneously co-align with the
magnetic field. In addition, the polar crystal structure of lacunar
spinels~\cite{Singh2014,Pocha2000} may be exploited for a
non-dissipative electric field control of the SkL.

\section*{Methods}

\textbf{Sample synthesis and characterization.} Single crystals of
GaV$_4$S$_8$ were grown by the chemical vapour transport method
using iodine as the transport agent. The crystals are typically
cuboids or hexagonal slabs with masses ranging from 1-60\,mg. The
sample quality was checked by powder X-ray diffraction, specific
heat and magnetization measurements. The crystallographic
orientation of the samples was determined by X-ray Laue and/or
neutron diffraction prior to the magnetization, SANS and AFM
studies. The magnetization measurements were performed using an MPMS
from Quantum Design. The field dependence of the magnetization was
measured in increasing temperature steps following an initial
zero-field cooling to 2\,K. The temperature dependence of the
magnetization was also measured in different fields. For both of
these approaches, we found no hysteresis in the magnetization curves
thus indicating that the SkL state cannot be quenched, i.e.
stabilized as a metastable state, in this compound.

\textbf{Non-contact atomic force microscopy.} All AFM data presented
were obtained by non-contact atomic force microscopy (nc-AFM)
performed in an Omicron cryogenic ultra-high vacuum (UHV) STM/AFM
instrument using the RHK R9-control electronics. The microscope
features interferometric detection which as a side-effect allows the
tuning of the cantilevers' effective quality factor $Q_{eff}$ by
varying the laser intensity used for the detection of the cantilever
motion. For all measurements we used magnetically coated AFM-tips
from the SSS-QMFMR series from NANOSENSORS\texttrademark in order to
achieve sensitivity to magnetic forces.

In nc-AFM mode the frequency $f$ of an oscillating cantilever is
measured via a phase locked loop and kept constant by the topography
feedback loop. The frequency shift $\Delta f$ with respect to the
resonance frequency $f_0$ far away from the sample  provides a
measure of the force gradient apparent between tip and sample.
Additionally, the oscillation amplitude $A$ is kept constant by
adjusting the excitation amplitude $A_{exc}$. Electrostatic
interactions between tip and sample have been minimized using an
additional FM-Kelvin control loop without any additional lateral
information in the measured contact potential difference.

We could not obtain magnetic contrast in magnetic force microscopy
mode, where the tip is retracted from the surface and the
conservative magnetic forces are measured via the induced frequency
shift $\Delta f$. Instead, a dissipative force interaction induced
the magnetic contrast in the excitation channel when the tip was
brought close to the sample surface. The nature of this dissipative
interaction is subject to further investigations. We could obtain
sufficient magnetic contrast only in images recorded below 10 K,
which is likely related to the low spin density of the material
corresponding to a single spin per formula unit and the strong
fluctuations of the S=1/2 spins.

With knowledge of the cantilever's spring constant $k$, the actual
oscillation amplitude $A$, the resonance frequency $f$, the
effective quality factor $Q_{eff,0}$ and the signal-to-drive ratio
$R_0$ far away from the sample we compute the dissipated power from
the measured excitation amplitude as
\begin{equation}
P = \frac{\pi k f R_0A}{Q_{eff,0}}  \cdot A_{exc}.
\end{equation}
For the images in figure 2 these parameters are $k=3.4$\,N$/$m,
$A=3$\,nm, $Q_{eff,0}=4.28\cdot10^5$, $f=70.87$\,kHz,
$R_0=30.92$\,$\mu$m$/$V with excitation amplitudes ranging from
150\,$\mu$V to 1.1\,mV.

\textbf{Small-angle neutron scattering.} Small-angle neutron
scattering (SANS) was used to study the long-wavelength, microscopic
magnetic states in a 25.5\,mg single crystal sample of GaV$_4$S$_8$.
The SANS measurements were carried out using the D33 instrument at
the Institut Laue-Langevin (ILL), Grenoble, France, and the SANS-II
instrument at the Swiss Spallation Neutron Source (SINQ), Paul
Scherrer Institut, Switzerland. In a typical instrument
configuration, neutrons of wavelength 5\,{\AA} were selected with a
FWHM spread ($\Delta\lambda$/$\lambda$) of ~10\%, and collimated
over a distance of 5.3\,m before the sample. The scattered neutrons
were collected by a two-dimensional multi-detector placed 5\,m
behind the sample.

The single crystal sample of GaV$_4$S$_8$ was mounted inside a
horizontal field cryomagnet that was installed on the beamline so
that the direction of applied magnetic field was approximately
parallel to the neutron beam. The sample was oriented to have a
horizontal scattering plane that included the [100] and [111]
crystal directions. Applying the magnetic field along either of
these directions was achieved by simply rotating the sample stick
inside the cryomagnet. The SANS measurements were done by both
tilting and rotating the sample and cryomagnet together through
angular ranges that moved the magnetic diffraction peaks through the
Ewald sphere. The SANS patterns presented in Fig.~3 were constructed
by summing together the detector measurements taken at each rotation
angle, which allows all of the Bragg spots to be presented in a
single image.

\textbf{Theoretical calculations.} The Monte-Carlo calculations for
Fig.~\ref{Fig4} are performed for the model in Eq.~\ref{eq:model} on
the two-dimensional triangular lattice with $144  \times 144$ sites,
while those for Fig.~\ref{Fig5} are for the model in
Eq.~\ref{eq:model} on the three-dimensional cubic lattice with $72
\times 72 \times 8$ sites. For both cases periodic boundary
conditions are imposed. The XXZ-type anisotropy in the exchange term
comes from the orbital ordering with $3z^2-r^2$-type orbitals
pointing in the [111] direction taken as the $z$ axis in the
calculation~\cite{Pocha2000}. The ratio of $J_{\parallel}/J_{\perp}$
was determined from a mean-field analysis of the low-temperature
magnetization curves, with the value of $D/J_{\perp}$ fixed
according to the observed cycloidal pitch, $a_{cyc}$. As long as the
magnetic states of slow spatial variation such as cycloid and
skyrmion are concerned, influence from complicated crystal structure
becomes reduced, which justifies the choice to study the phase
diagram using a lattice spin model on the simple cubic lattice. In
the latter case, the Dzyaloshinskii-Moriya term is given by
\begin{eqnarray}
\mathcal{H}_{\rm DM}=-D \sum_{i} \left[
 \hat{\bm y} \cdot (\bm m_i \times \bm m_{i+\hat{x}})
-\hat{\bm x} \cdot (\bm m_i \times \bm m_{i+\hat{y}}) \right].
\end{eqnarray}

By comparing the energies of the three magnetic phases, we find that
only the cycloidal and the ferromagnetic states are stable at $T$=0,
while the SkL state is not. The transition points at finite $T$ are
determined according to the calculated temperature-dependence of the
specific heat and the net magnetization. The specific heat has a
peak at the transition from the paramagnetic to the ordered phases.
At the transition from the ferromagnetic phase to the cycloidal or
the SkL phase, the magnetization shows a sharp drop with increasing
temperature, while the specific heat shows almost no anomaly.
Magnetic structures in these phases are identified from both the
real-space and the reciprocal space mappings of their local
magnetization configurations.

\textbf{Acknowledgements} This work was supported by the Hungarian
Research Funds OTKA K 108918, OTKA PD 111756 and Bolyai 00565/14/11,
by the European Research Council Project CONQUEST, by the Swiss NSF
Grant No. 200021\_153451, by the DFG under grant no. SFB 1143 and
via the Transregional Research Collaboration TRR 80 From Electronic
Correlations to Functionality (Augsburg/Munich/Stuttgart) and by
JSPS KAKENHI under Grant Nos. 25870169 and 25287088 from MEXT Japan.

\textbf{Author Contributions} I.K., S.B., P.M., E.N., L.M.E.,
J.S.W., C.D.D., D.E., V.T. performed the measurements; I.K., S.B.,
P.M., E.N., H.M.R., J.S.W., A.L. analysed the data; V.T., H.N.
contributed to the sample preparation; M.M., K.Y. developed the
theory; I.K. wrote the manuscript and planned the project.

\textbf{Additional information} The authors declare no competing
financial interests.

\end{document}